\documentclass[a4paper,12pt]{article}

\usepackage{graphicx}

\begin{document}

\begin{center}
{\large\bf b-s anomaly decays in covariant quark model}

{Aidos Issadykov$^{1,2,\P}$}, {Mikhail A.~Ivanov$^{1}$}

$^1${Joint Institute for Nuclear Research, 141980, Dubna, Russia}

$^2${The Institute of Nuclear Physics,Ministry of Energy of the Republic of Kazakhstan, Almaty,Kazakhstan}

$^\P${E-mail: issadykov@jinr.ru}
\end{center}

\centerline{\bf Abstract}

The work is devoted to the study of b-s anomaly decays.
We evaluated branching fractions of $B\to K^\ast \mu^+\mu^-$,  $B^0_s\to \phi\mu^+\mu^-$ and   $B_s\to \mu^+\mu^-$ decays and compared them with 
available experimental data and with results from other theoretical approaches.

\section{Introduction}
Worth noting that Standard model has been extremely successful in explaining
the results of experiments for particle physics. The outstanding success of SM in the description of almost all experimental data in particle physics is
manifested in the electroweak pool for different observables.
Nevertheless, in recent years observed discrepancies in $B$-meson rare decays
with the predictions of SM. Flavour-changing neutral currents have been
prominent tools in high-energy physics in the search for new degrees of freedom, due
to their quantum sensitivity to energies much higher than the external particles
involved and can be instrumental in order to determine where to look for new
physics.
During the last decade a lot of observables, including the branching ratios, CP
and the angular asymmetry in inclusive and exclusive decay modes of $B$-meson were
measured by B-factories and at LHC experiments. These data allow to explore the
spiral structure in the interactions with the flavour-changing and a possible existence
of new sources of CP violation.

In 2013 it has been paid much attention
to the rare flavor-changing neutral current decay 
$B\to K^\ast (\to K\pi)\mu^+\mu^-$. One of the reason was the first measurement of
form-factor independent angular observables 
performed by LHCb Collaboration \cite{Aaij:2013qta,Aaij:2013iag}. 
It has been claimed that there is a 3.7$\sigma$ deviation from the Standard Model for so-called $P^{\prime}_{5}$ angular obsevable.

The $B^0_s\to \phi\ell^+\ell^-$ decay is similar to the $B\to K^\ast\ell^+\ell^-$ decay.
The $B_s$ meson production is suppressed compared to the  $B^0$ meson by the relation $f_s/f_d \simeq 1/4$, but
the narrow resonance $\phi$ provides a clean set of data with low background.
The main difference between $ B^0_s\to \phi\ell^+\ell^-$ and $B\to K^\ast\ell^+\ell^-$
decays is that the final state do not contain information about the initial state
of the meson, whether it was $B_s$ or $\bar{B_s}$. 
$B^0_s\to \phi\ell^+\ell^-$ decay channel was first discovered and studied by CDF Collaboration in 2011\cite{Aaltonen:2011qs,Aaltonen:2011cn},
later been studied by LHCB Collaboration \cite{Altmannshofer:2014rta,Aaij:2013aln}.
Despite the fact that the angular distributions are in good
agreement with the SM expectations,
branching ratio of decay had a $3.1 \sigma$ disagreement with the prediction
of the SM \cite{Altmannshofer:2014rta,Lyon:2014hpa}.

Our goal was to check all this deviations of theoretical predictions taken into account the last experimental data.

\section{The $B\to K^\ast (\to K\pi)\ell^+\ell^-$ decay}

The study of the rare $B\to K^\ast \ell^+\ell^-$~decay in the framework 
of the covariant quark model with infrared confinement was made in  \cite{Dubnicka:2015iwg}. The obtained results compared with 
available experimental data and the results from other theoretical approaches.
We give in Table~\ref{tab:branching}  the numerical value for
the total branching ratio ${\cal B} (B \to K^{*} \mu^{+} \mu^{-})$ and compare them with available experimental data and with other approaches.
\begin{table}[ht]
\begin{center}
\caption{Total branching fractions (CQM - covariant quark model).}
\begin{tabular}{c|cccc}
\hline
Mode 					& CQM				& \multicolumn{2}{c}{Others} 					& Expt.~\cite{Agashe:2014kda,Aaij:2014pli,Pescatore:2014ksa} \\
\hline
$B\to K^\ast\mu^+\mu^- $ 	& $12.7\times 10^{-7}$ 		& $(11.9 \pm 3.9) \times10^{-7}$ &\cite{Ali:2002jg} 	&$(9.24\pm 0.93 ({\rm stat}) \pm 0.67 ({\rm sys}))\times 10^{-7}$ \\
& & $11.5 \times 10^{-7}$ &\cite{Melikhov:1997wp}& \\
& & $14 \times 10^{-7}$ &\cite{Geng:1996az}& \\
\hline
\end{tabular}
\label{tab:branching}
\end{center}
\end{table}


Also we explored the influence of the intermediate scalar $K^\ast_0$
meson on the angular decay distribution of the cascade decay
$B\to K\pi+\mu^+\mu^-$ in \cite{Issadykov:2015iba}.  In the wake of exploring uncertainty in the full angular distribution of the $B\to K\pi+\mu^+\mu^-$ decay caused
by the presence of the intermediate scalar $K^\ast_0$ meson. We give in Table~\ref{tab:Kstar}  the numerical values for
the total branching ratios of $(B\to K^\ast_0(800)\ell^+\ell^-)$ and $B^0_d\to K_0^{\ast\,0}(800) \bar\nu\nu$  decays and compare  with another theoretical prediction.

\begin{table*}
\begin{center}
\def\arraystretch{1.5}
{\begin{tabular}{|l|c|c|}
\hline 
Decay modes  &  \multicolumn{2}{c|}{Branching fractions} \\
\cline{2-3} & CQM    & \cite{Wang:2014upa}\\
\hline
$B_d^0\to  K_0^{\ast\,0}(800) \mu^+\mu^-$   	&$3.47\times 10^{-7}$   	&  $(7.31\pm 1.21)\times 10^{-7}$\\

$B_d^0\to  K_0^{\ast\,0}(800) \tau^+\tau^-$ 	& $0.61\times 10^{-7}$  	& $(1.33\pm 0.36)\times 10^{-7}$\\      

$B^0_d\to K_0^{\ast\,0}(800) \bar\nu\nu$ 		&$2.53\times 10^{-6}$  	& $(6.30\pm 0.97)\times 10^{-6}$\\
\hline
\end{tabular}
\caption{ The branching fraction for  $(B\to K^\ast_0(800)\ell^+\ell^-)$ and $B^0_d\to K_0^{\ast\,0}(800) \bar\nu\nu$  decays.}
\label{tab:Kstar}}
\end{center}
\end{table*}

Let us briefly discuss the impact of scalar resonance $K^\ast_0$
on $B\to K^\ast(\to K\pi)\ell^+\ell^-$ decay. As well known,
the narrow $K^\ast(892)$ vector resonance is described by 
a Breit-Wigner parametrization and the given cascade B-decay can be calculated
by using the narrow width approximation. But it is not  true in the case
of the broad scalar $K^\ast_0(800)$ meson. We will use for the time
being the parametrization accepted in Ref.~\cite{Meissner:2013pba} 
which integrated value in the $K^\ast$-resonance region is equal to
\begin{equation}
\int_{ (m_{K^\ast} - \delta_m)^2 }^{ (m_{K^\ast}+\delta_m)^2 } dm_{K\pi}^2 |L_S(m_{K\pi}^2)|^2 = 0.17,  \quad \delta_m = 100 MeV.
\label{eq:scale}
\end{equation}
Then we scale the calculated value for the differential decay rate 
$d\Gamma(B\to K^\ast_0(800)\mu^+\mu^-)$ by this factor and compare
with those for $B\to K(892)\mu^+\mu^-$ decay. 
The integrated ratio
\begin{equation}
R(q^2)= \frac{2/3\, d\Gamma(B\to K^\ast(892)\mu^+\mu^-)} {2/3\, d\Gamma(B\to K^\ast(892)\mu^+\mu^-) + 0.17 d\Gamma(B\to K^\ast_0(800)\mu^+\mu^-)}
\end{equation}
(numerator and denominator are integrated separately
in the full kinematical region of $q^2$ $\!$)  gives
the size of the S-wave pollution to the
branching ratio of the $B\to K^\ast\ell^+\ell^-$ decay only
about 6$\%$. The S-wave contribution near threshold was disscussed in Ref.~\cite{Kang:2013jaa}

\section{The $ B^0_s\to \phi(\to K^+K^-)\ell^+\ell^-$ decay}

In paper\cite{Dubnicka:2016nyy} we calculated all form factors which appear
in the $B_s\to\phi$~transition.
The expressions for the Wilson coefficients $C_7$ and $C_9$  
are taken on the two-loop level of accuracy by using the results
obtained in Refs.~\cite{Asatryan:2001zw,Greub:2008cy}.
Then we evaluated the branching fraction, the forward-backward 
asymmetry and the so-called optimized observables using form factors in the cascade decay
$B_s\to \phi(\to K^+K^-)\mu^+\mu^-$.  We compared our results 
with the recent experimental data reported in Ref.~\cite{Aaij:2015esa}
for various $q^2$-bins. 

\begin{footnotesize}
\begin{table}
\caption{Binned observables for$(B_s\to \phi\mu^+\mu^-)$ decay.}
\label{tab:bin}
\begin{tabular}{|cllll|}
\hline
$ 10^7 {\cal B}(B_s\to \phi\mu^+\mu^-) $  & 2 loop(CQM)  & 1 loop(CQM)  & \cite{Descotes-Genon:2015uva} & Expt.~\cite{Aaij:2015esa}\\
\hline
 $[0.1,2]$ & $0.99\pm 0.2$   & $0.86\pm 0.17 $ 
& $1.81 \pm 0.36$ & $1.11 \pm 0.16$  \\

 $[2,5]$   & $0.90\pm 0.18$  & $0.95\pm 0.19 $  
& $1.88\pm 0.31$  & $0.77\pm 0.14$   \\

 $[5,8]$   &  $--$           & $1.25\pm 0.25$
&  $2.25\pm 0.41$ & $0.96\pm 0.15$   \\

 $ [11,12.5]$ & $0.84\pm 0.17$&$0.88\pm 0.18$
& $--$ & $ 0.71 \pm 0.12 $  \\

 $ [15,17]$ & $1.15\pm 0.23$&$1.19\pm 0.24$
& $--$ & $ 0.90 \pm 0.13 $  \\

 $ [17,19]$ & $0.75\pm 0.15$&$0.77\pm 0.15$
& $--$ & $ 0.75 \pm 0.13 $ \\

 $ [1.,6.]$ & $1.56\pm 0.31$&$1.64\pm 0.33$
& $--$ & $ 1.29 \pm 0.19 $ \\

 $[15,19]$ & $1.89\pm 0.28$  & $1.95\pm 0.29$ 
& $2.20\pm 0.16$  & $1.62\pm 0.20$   \\
\hline
\hline
$ F_L (B_s\to \phi\mu^+\mu^-) $  & 2 loop(CQM) & 1 loop(CQM) 
&\cite{Descotes-Genon:2015uva} & Expt.~\cite{Aaij:2015esa} \\ 
\hline
 $[0.1,2]$ & $0.37\pm 0.04$  & $0.46\pm 0.05$     
         & $0.46\pm 0.09$  & $0.20\pm 0.09$   \\ 

 $[2,5]$   & $0.72\pm 0.07$  & $0.74\pm 0.07$
         & $0.79\pm 0.03$  & $0.68\pm 0.15$   \\ 
 
  $ [1,6] $ 	& $0.69\pm 0.07$&$0.71\pm 0.07$
 				& $--$ & $ 0.63 \pm 0.09 $  \\ 

 $[15,19]$ & $0.34\pm 0.03$  & $0.34\pm 0.03$ 
         & $0.36\pm 0.02$  & $0.29\pm 0.07$ \\ 
\hline
\end{tabular}
\end{table}
\end{footnotesize}

The level of agreement with experiment can be estimated by combining in 
quadrature the experimental errors with the theoretical ones: if the 
difference in observable values is smaller, then it can be seen as 
compatible with zero.

Using this optics one can address the $3.3 \sigma$ deviation seen by 
\cite{Aaij:2015esa} for branching fraction in the $1-6$ GeV range. 
In the covariant confined quark model this discrepancy is much reduced. 
The remaining deviation ($  1.4 \sigma$) shrinks is even further if 
the two-loop corrections for the Wilson coefficients are taken into account, 
down to $  1.1 \sigma$. With such error reduction one cannot claim a 
discrepancy with the SM any longer.

Overall one observes a good description of the data by the covariant quark 
model and the agreement becomes even better if the two-loop corrections are 
taken into account. The biggest discrepancy of $  2.5 \sigma$ observed for 
$F_L$ in the lowest bin $0.1 \le q^2 \le 2$ GeV is reduced to $  1.7 \sigma$ 
when these corrections are taken into account. 

\section{The $ B^0_s\to \ell^+\ell^-$ decay}
As it was shown in \cite{Buchalla:1995vs} the rare decays are fully dominated
by internal top quark contributions. We used next definition for the branching ratio of $ B^0_s\to \ell^+\ell^-$ to check the sensivity to the top quark mass \cite{Buchalla:1995vs}:
\begin{equation}\label{bbll}
B(B_s\to l^+l^-)=\tau(B_s)\frac{G^2_F}{\pi}
\left(\frac{\alpha}{4\pi\sin^2\Theta_W}\right)^2 F^2_{B_s}m^2_l m_{B_s}
\sqrt{1-4\frac{m^2_l}{m^2_{B_s}}} |V^\ast_{tb}V_{ts}|^2 Y^2(x_t)
\end{equation}
where $B_s$ denotes the flavor eigenstate $(\bar bs)$ and $F_{B_s}$ is
the corresponding decay constant.
The function $Y(x_t)$ is given by 
\begin{equation}
Y(x_t)= \eta_Y Y_0(x_t)
\end{equation}
where 
\begin{equation}
Y_0(x_t)= \frac{x}{8} (\frac{4-x}{1-x}+\frac{3x}{(1-x)^2}\ln{x}), \quad \quad x=m^2/M^2_W,\quad \quad \eta_Y=1.028 .
\end{equation}
Thereotical predictions within the covariant quark model for the branching ratio of $ B^0_s\to \ell^+\ell^-$ decay wtih top quark contribution given in Table~\ref{tab:bmumu}.

\begin{table}[t]
\caption{\label{tab:bmumu}
         $B_s$ meson weak decays.}
\def\arraystretch{1.5}
\begin{center}
\begin{tabular}{|l|l|l|l|}
\hline
 Mode & CQM & Experiment \cite{Aaij:2017vad},\cite{Aaij:2017xqt} &SM expectaion\\
\hline
$B_s^- \to \mu \mu$      	&$4.31*10^{-9} $		&$(3.0 \pm0.6 ^+0.3 _-0.2)*10^{-9}$	&$(3.1*10^{-9} <Br > 	5.0*10^{-9})$\\

$B_s^- \to \tau \tau$      	&$0.92*10^{-6}$		&$<0.68*10^{-4}$									&$(0.77 \pm 0.05)*10^{-6}$\\
\hline
\end{tabular}
\end{center}
\end{table}

\section*{Acknowledgment}
Authors A. Issadykov and M.A. Ivanov acknowledge the partial support by the
Ministry of Education and Science of the Republic of Kazakhstan, grant 3092/GF4, state registration No. 0115RK01040. 

Author A. Issadykov is grateful for the support by the JINR, grant number 17-302-03.


\begin{thebibliography}{99}

\bibitem{Aaij:2013qta}
  R.~Aaij {\it et al.} [LHCb Collaboration],
  Phys.\ Rev.\ Lett.\  {\bf 111} (2013) 191801
  doi:10.1103/PhysRevLett.111.191801
  [arXiv:1308.1707 [hep-ex]].

\bibitem{Aaij:2013iag}
  R.~Aaij {\it et al.} [LHCb Collaboration],
  JHEP {\bf 1308} (2013) 131
  doi:10.1007/JHEP08(2013)131
  [arXiv:1304.6325 [hep-ex]].

\bibitem{Aaltonen:2011qs}
  T.~Aaltonen {\it et al.} [CDF Collaboration],
  Phys.\ Rev.\ Lett.\  {\bf 107} (2011) 201802
  doi:10.1103/PhysRevLett.107.201802
  [arXiv:1107.3753 [hep-ex]].

\bibitem{Aaltonen:2011cn}
  T.~Aaltonen {\it et al.} [CDF Collaboration],
  Phys.\ Rev.\ Lett.\  {\bf 106} (2011) 161801
  doi:10.1103/PhysRevLett.106.161801
  [arXiv:1101.1028 [hep-ex]].

\bibitem{Altmannshofer:2014rta}
  W.~Altmannshofer and D.~M.~Straub,
  Eur.\ Phys.\ J.\ C {\bf 75} (2015) no.8,  382
  doi:10.1140/epjc/s10052-015-3602-7
  [arXiv:1411.3161 [hep-ph]].

\bibitem{Aaij:2013aln}
  R.~Aaij {\it et al.} [LHCb Collaboration],
  JHEP {\bf 1307} (2013) 084
  doi:10.1007/JHEP07(2013)084
  [arXiv:1305.2168 [hep-ex]].

\bibitem{Lyon:2014hpa}
  J.~Lyon and R.~Zwicky,
  arXiv:1406.0566 [hep-ph].

\bibitem{Dubnicka:2015iwg}
  S.~Dubni\v{c}ka, A.~Z.~Dubni\v{c}kov\'{a}, N.~Habyl, M.~A.~Ivanov, A.~Liptaj and G.~S.~Nurbakova,
  Few Body Syst.\  {\bf 57} (2016) no.2,  121
  doi:10.1007/s00601-015-1034-4
  [arXiv:1511.04887 [hep-ph]].

\bibitem{Agashe:2014kda}
  K.~A.~Olive {\it et al.} [Particle Data Group],
  Chin.\ Phys.\ C {\bf 38} (2014) 090001.

\bibitem{Aaij:2014pli}
  R.~Aaij {\it et al.} [LHCb Collaboration],
  JHEP {\bf 1406} (2014) 133
  doi:10.1007/JHEP06(2014)133
  [arXiv:1403.8044 [hep-ex]].

\bibitem{Pescatore:2014ksa}
  L.~Pescatore [LHCb Collaboration],
  arXiv:1410.2411 [hep-ex].

\bibitem{Ali:2002jg}
  A.~Ali, E.~Lunghi, C.~Greub and G.~Hiller,
  Phys.\ Rev.\ D {\bf 66} (2002) 034002
  doi:10.1103/PhysRevD.66.034002
  [hep-ph/0112300].

\bibitem{Melikhov:1997wp}
  D.~Melikhov, N.~Nikitin and S.~Simula,
  Phys.\ Rev.\ D {\bf 57} (1998) 6814
  doi:10.1103/PhysRevD.57.6814
  [hep-ph/9711362].

\bibitem{Geng:1996az}
  C.~Q.~Geng and C.~P.~Kao,
  Phys.\ Rev.\ D {\bf 54} (1996) 5636
  doi:10.1103/PhysRevD.54.5636
  [hep-ph/9608466].

\bibitem{Issadykov:2015iba}
  A.~Issadykov, M.~A.~Ivanov and S.~K.~Sakhiyev,
  Phys.\ Rev.\ D {\bf 91} (2015) no.7,  074007
  doi:10.1103/PhysRevD.91.074007
  [arXiv:1502.05280 [hep-ph]].

\bibitem{Wang:2014upa} 
  Z.~G.~Wang,
  ``Semi-leptonic $B\to S$ decays in the standard model and 
    in the universal extra dimension model,''
    arXiv:1411.7961 [hep-ph].

\bibitem{Meissner:2013pba}
  U.~G.~Mei{\ss}ner and W.~Wang,
  JHEP {\bf 1401} (2014) 107
  doi:10.1007/JHEP01(2014)107
  [arXiv:1311.5420 [hep-ph]].

\bibitem{Kang:2013jaa}
  X.~W.~Kang, B.~Kubis, C.~Hanhart and U.~G.~Meißner,
  Phys.\ Rev.\ D {\bf 89} (2014) 053015
  doi:10.1103/PhysRevD.89.053015
  [arXiv:1312.1193 [hep-ph]].

\bibitem{Dubnicka:2016nyy}
  S.~Dubni\v{c}ka, A.~Z.~Dubni\v{c}kov\'{a}, A.~Issadykov, M.~A.~Ivanov, A.~Liptaj and S.~K.~Sakhiyev,
  Phys.\ Rev.\ D {\bf 93} (2016) no.9,  094022
  doi:10.1103/PhysRevD.93.094022
  [arXiv:1602.07864 [hep-ph]].


\bibitem{Asatryan:2001zw}
  H.~H.~Asatryan, H.~M.~Asatrian, C.~Greub and M.~Walker,
  Phys.\ Rev.\ D {\bf 65} (2002) 074004
  doi:10.1103/PhysRevD.65.074004
  [hep-ph/0109140].

\bibitem{Greub:2008cy}
  C.~Greub, V.~Pilipp and C.~Schupbach,
  JHEP {\bf 0812} (2008) 040
  doi:10.1088/1126-6708/2008/12/040
  [arXiv:0810.4077 [hep-ph]].

\bibitem{Aaij:2015esa}
  R.~Aaij {\it et al.} [LHCb Collaboration],
  JHEP {\bf 1509} (2015) 179
  doi:10.1007/JHEP09(2015)179
  [arXiv:1506.08777 [hep-ex]].

\bibitem{Descotes-Genon:2015uva}
  S.~Descotes-Genon, L.~Hofer, J.~Matias and J.~Virto,
  JHEP {\bf 1606} (2016) 092
  doi:10.1007/JHEP06(2016)092
  [arXiv:1510.04239 [hep-ph]].

\bibitem{Buchalla:1995vs}
  G.~Buchalla, A.~J.~Buras and M.~E.~Lautenbacher,
  Rev.\ Mod.\ Phys.\  {\bf 68} (1996) 1125
  doi:10.1103/RevModPhys.68.1125
  [hep-ph/9512380].

\bibitem{Aaij:2017vad}
  R.~Aaij {\it et al.} [LHCb Collaboration],
  Phys.\ Rev.\ Lett.\  {\bf 118} (2017) no.19,  191801
  doi:10.1103/PhysRevLett.118.191801
  [arXiv:1703.05747 [hep-ex]].
  
\bibitem{Aaij:2017xqt}
  R.~Aaij {\it et al.} [LHCb Collaboration],
  arXiv:1703.02508 [hep-ex].

\end{thebibliography}
\end{document}